\documentclass{llncs}
\usepackage{graphicx}
\usepackage[caption=false]{subfig}
\usepackage{hyperref}
\usepackage{booktabs}
\usepackage{amsfonts}
\usepackage{booktabs}
\usepackage{siunitx}
\usepackage{multirow} 
\begin{document}
\title{PolicySpace: a modeling platform}
\titlerunning{PolicySpace}  
\author{Bernardo Alves Furtado\inst{1,2}}
\authorrunning{Furtado, B.} 
\tocauthor{Bernardo Alves Furtado}
\institute{Institute for Applied Economic Research (IPEA), Brazil\\
\email{bernardo.furtado at ipea.gov.br},\\ 
\and
National Council of Research (CNPq)}
\maketitle              
\begin{abstract}
Public Policy involves proposing changes to existing practices, alternatives, new habits. Citizens and institutions react accordingly, accepting, refuting or adapting. Agent-based modeling is a tool that can enrich the policy analysis package explicitly considering dynamics, space and individual-level interactions. This paper presents a modeling platform called PolicySpace that models public policies within an empirical, spatial environment using data from 46 metropolitan regions in Brazil. We describe the basics of the model, its agents and markets, the tax scheme, the parametrization, and how to run the model. Finally, we validate the model and demonstrate an application of the fiscal analysis. Besides providing the basics of the platform, our results indicate the relevance of the rules of taxes transfer for cities' quality of life. 
\end{abstract}

\section{Introduction and literature}
This paper conceptualizes, justifies and validates a policy-modeling platform: PolicySpace, that is available for download and immediate application. As such, the paper focus on the presenting the possibilities and potentialities of the platform itself, detailing how to run and test simulations with an added brief application of fiscal analysis \footnote{The platform is thoroughly detailed in \cite{furtado_policyspace:_2018}}

Public policy diagnostics, proposals, implementation and evaluation are complex tasks. They involve a number of different groups and interests, over idiosyncratic, spatial regions with consequences through varying time-span. As a norm, a given policy proposal will affect citizens, institutions and firms; cities and provinces, immediately or in the long-range with a multitude of levels of magnitude. 
    
Complex systems in turn focus on a high number of agents that interact over time and space providing emerging, nonlinear results. Thus, complex systems’ methodologies are designed to handle exactly such interactions and the unfolding and unraveling of such systems. 
    
Agent-based modeling (ABM) is one of the methodologies of complex systems that may shed some light into policy comprehension. Flexible and cost-effective, ABM helps delineate \textit{ex-ante} policy analysis that anticipates effects evaluate alternatives, working very much as experiments, which are costly and often impractical within the social sciences realm.
    
Applications of ABMs for social science have been around for some time \cite{epstein_growing_1996,schelling_process_1972}. However, recent handbooks have further defined the concepts and applications \cite{wilensky_introduction_2015} and evaluated models for specific domains of social sciences, from economics \cite{boero_agent-based_2015} and social science in general \cite{edmonds_simulating_2013,helbing_social_2012} to political science \cite{geyer_handbook_2015}, geography \cite{heppenstall_agent-based_2012} and spatial analysis \cite{batty_generic_2012}.

All of them consider ABM as a computational artificial tool that provides an environment in which agents (of all sizes and shapes) interact among themselves and with the environment in a continuous time-driven direction. On top of such artificial construct, all kinds of experimentation can be made with considerably easy changes and tweaks using a relatively wide range of possible tools \cite{abar_agent_2017}.

Applications of policy are relatively fewer, but have been tried in transportation \cite{erath_large-scale_2012} and economics \cite{baptista_macroprudential_2016}. Mostly, applications have been praised as likely potential applications \cite{colander_complexity_2014,oecd_applications_2009} or they have been used for specific markets and context, such as the labor market \cite{neugart_agent-based_2012}, financial analysis and interbank dependence \cite{tabak_topological_2009} or macroeconomics \cite{dosi_schumpeter_2010}. Comprehensive models of the whole economy are typically simple \cite{gaffeo_adaptive_2008,lengnick_agent-based_2013} or too complex \cite{dawid_agent-based_2014}. 

PolicySpace is an open AB model that has been designed and customized for the analysis of Brazilian metropolitan regions, more precisely, Areas of Concentrated Population (ACPs) \footnote{ACPs are called Functional Urban Areas by the OECD \cite{ahrend_what_2014}.}. PolicySpace includes citizens, families, firms and municipalities' government that interact in three markets: goods, labor and real estate. It also enables fiscal analysis (via implementation of five different taxes), territorial experimentation as well as tweaking of 20 different parameters that govern behaviors in the firms, families, government and the markets. Finally, PolicySpace has been through extensive validation and proved itself to be robust both to the variation of parameters, rules or ACPs whilst simultaneously providing useful insights. Nevertheless, the modularity of PolicySpace and its continuous in-development status certainly provides room for improvements.  

PolicySpace is intended as a simple model that covers the economy but that also offers alternative analysis for urban mobility, territorial fiscal analysis and real estate understanding, all in an interconnected manner. Although PolicySpace departs from Gaffeo \cite{gaffeo_adaptive_2008} and Lengnick \cite{lengnick_agent-based_2013}, and then from an abstract earlier version \cite{furtado_simple_2016}, we have no information of other AB models that simultaneously cover municipalities, economic markets and agents as presented here. Specifically, the contribution of this paper is to provide an overview to the platform that enables the interested reader to its understanding, usefulness and usability. Besides this introduction, the text describes the platform and its validation and concludes with a list of alternatives for further developing policy applications. 

\section{PolicySpace}
The description of the model in \cite{furtado_policyspace:_2018} follows the ODD protocol \cite{grimm_odd_2010}. Here, we describe the intuition, the entities and the schedule, providing a general idea of the sub-models. We then go on to detail the parameters and how to run the model.

\subsection{Intuition}
The platform mimics the politically administrative municipalities as geographically spatial and fiscal entities. Families and firms are the main agents that interact among themselves. This configuration of municipalities, families and firms enables the labor market and a goods market along with the collection of three relatively large taxes: on labor via salaries, on consumption via sales and on production and profits via firms. Additionally, a real estate market provides further interaction among families that are mobile and free to move about among the municipalities. The real estate market then enables the tax on properties via residents and taxes on estate transactions whenever they happen. 

\subsection{Agents}
PolicySpace is built in Python and follows the Object-Oriented Programming paradigm. As such, there are five classes of agents. Citizens, families, residences, firms and municipality governments. The proportions of each agent is based on actual data for the year 2000. 
	
Citizens work and commute individually. However, their consumption, housing and saving decisions are made together as a family. Citizens have distinct ages, qualification and gender. They are important, respectively, to enter the labor market, to qualify its level of production and to allow for birth giving. In the current configuration, the only attribute that changes dynamically is the agents’ ages. New citizens are ‘born’ following official fertility rates for female agents and are incorporated automatically to the mothers’ family. Also according to official data, citizens of all ages pass away, given their ages, state of residence and gender. 

Families are mobile moving occasionally from house to house depending on their savings and employment capabilities. Houses are fixed throughout the simulation with a given quality, surface and location. However, house prices vary depending on the success and rate of their municipalities’ ability to collect taxes and invest them, given their population at each month.

Firms are also fixed and engage in monthly production, which is directly proportional to the number of employees and their qualification \cite{gaffeo_adaptive_2008}. Firms observe their stock in order to decide on prices \cite{hamill_agent-based_2016,seppecher_what_2017}, their revenue and general unemployment to decide on wages \cite{dosi_schumpeter_2010} and  their profits to decide on hiring and firing \cite{neugart_agent-based_2012}.

Finally, municipalities are spatial entities \cite{grimm_odd_2010} that basically manage taxes collected within their boundaries and invest them into improved quality of life weighted by their population. 

\subsection{Schedule and markets}
Before the simulation, all agents described above are generated provided that a population percentage, an average family size and an estimate of house vacancy and the official data input have been set. Before the actual simulation starts, a labor market round is set in motion so that the level of unemployment for January 2000 is achieved. Henceforth, the simulation runs on a monthly and sequential basis.
\begin{enumerate}
\item Production is deterministically calculated from the number of employees for each firm and their level qualification and a productivity parameter alpha. 
\item Demographics is then processed with agents aging, dying or having children.
\item Next, families go out to consume. Firstly, families decide on how much to consume, following a beta parameter on propensity to consume. Money not set to consume in a given month is saved without liquidity to be later invested in the real estate market. Families decide on a sample of firms to consume either by cheapest price or closest firm. Firms offer the amount of goods demanded by each family on a first come first served basis as long as they have goods in stock. Companies collect taxes on consumption.
\item Once consumption has finished, firms have all elements to decide on prices, salaries and firing or hiring. 
\item The labor market follows with all unemployed agents of age applying and all firms that have profits hiring. Firms paying higher salaries choose first. They may hire the best qualified of the pool of candidates or the candidate that lives the closest, following a parameter. \footnote{That decision follows observed patterns for the Brazilian market in which the cost of transport is partially a responsibility of the company. It has also been proved robust in a sensitivity analysis \cite{furtado_policyspace:_2018}}  
\item In the real estate market, vacant houses are always on the market, whereas families enter the market occasionally, following a set parameter. The simulation always has more houses than families and thus some families own more than one house. The family with the highest savings make the first bid on the demand side. Offering prices are hedonically calculated for each house, given their quality, location and surface. The matching is made sequentially with transaction prices being averaged between bid and offering. Estate transaction taxes are collected at the moment of the sale. 
\item Once all markets have run, the taxes distribution takes effect and each municipality invests their quote directly on quality of life. 
\item All data is saved and a new month begins until the number of months set by the modeler. 
\end{enumerate}
The model is set to start in the year 2000 and run for a maximum of 30 years, constrained by available data projections. However, typical runs are for 20 years (or 5,040 working days). 

\subsection{Taxes analysis}

PolicySpace contains four alternative ways to distribute the taxes collected. They are implemented using two parameters: \texttt{Alternative0} and \texttt{FPM\_distribution}. The default configuration has both of them set as $true$. \texttt{Alternative0} refers to the fact that municipalities are autonomous and receive taxes according to the rules in place. When $false$, this parameter implies that all of the municipalities of a given ACP are considered as one for fiscal distribution purposes. This hypothesis is supported by the fact that larger municipalities are more efficient in their spending \cite{gasparini_transferencias_2011} and receive a larger proportion of taxes per capita \cite{furtado_fatos_2013}. Further, less fragmentation of municipalities leads to more productivity \cite{ahrend_what_2014}. \texttt{FPM\_distribution} in turn sets the rules of FPM (a municipal tax transfer) in effect or not. The combination of the four alternatives and the criteria for distribution along with the allocated percentage of taxes can be seen from \ref{tab1}.
\begin{table}
\centering
\begin{tabular}{l|ccc|cc|c|c} 
\toprule
{Options} & & {Alt0:True} & & {Alt0:False} & & {Alt0:True} & {Alt0:False} \\
{} & & {FPM:True} & & {FPM:True} & & {FPM:False} & {FPM:False}\\
\midrule
{Taxes} & {Mun.} & {State} & {FPM} & {State} & {FPM} & {Mun.} & {State} \\
\midrule
{Consumption} & .1875 & .8125 & & 1 & & 1 & 1 \\
{Labor} & & .765 & .235 & .765 & .235 & 1 & 1 \\
{Transaction} & 1 & & & 1 & & 1 & 1 \\
{Firms} & & .765 & .235 & .765 & .235 & 1 & 1 \\
{Property} & 1 & & & 1 & & 1 & 1 \\
\midrule
{Criteria} & Locally & Equally & FPM & Equally & FPM & Locally & Equally \\
\bottomrule
\end{tabular}
\caption{Alternative distributions of collected taxes among municipalities. Depending on the set of rules, taxes are distributed equally among all municipalities, locally, so that a municipality receives only what was collected within the own municipality, according to FPM transfer rules, or a combination of them.}
\label{tab1}
\end{table}
\subsection{Running PolicySpace}
As the idea of the paper is also to present the platform, we describe the basics of running PolicySpace \href{https://github.com/bafurtado/policyspace}{GitHub.com/BAFurtado/PolicySpace}. Once installed, you can run the web version with the command: \texttt{python main.py web} and open the model in a browser: \texttt{http://127.0.0.1:5000/}. 
\begin{enumerate}
\item Run type: four alternatives are available:
\begin{enumerate}
\item The first: `run' is the running of a single model, for a single ACP, a single configuration and plotting and saving the results. 
\item The option `sensitivity' automatically varies any of the parameters. If they are booleans, only the name of the parameter will suffice. However, when the parameter is a quantity, the modeler needs to set the first value, the last value, and the number of times the parameter is going to be divided, separated by colons, as such: \texttt{python main.py sensitivity ALPHA:.04:.94:7}. 
\item The `distributions' option tests and plots the four automatic alternatives of taxes distributions together. 
\item Finally, the `acps' type runs the model for each one of the 46 ACPs available in the model, at least once for each one.
\end{enumerate}
\item Number of runs per config: on top of any of the runs selected in the first option, the modeler may set the number of runs for each configuration in order to consider stochasticity. For example, if the modeler is running a sensitivity analysis on \texttt{ALPHA}, for seven different values, then she may choose to run 4 times for each value. That means that PolicySpace will run 28 times for this example. The output is organized as such that the plots and files will manage averages and folders.
\item Number of cores to use: is a typically operational decision. The default $(-1)$ is to run on all cores of your machine.
\item Here we discuss the main parameters briefly:
\begin{enumerate}
\item \texttt{ALPHA}: is an exponent applied to the years of qualification of employees in order to qualify	productivity.
\item \texttt{ALTERNATIVE0} takes `false' or `true' and determines whether municipalities function as they are presently (true) or as if they are together, as a single unit for fiscal purposes. 
\item \texttt{BETA}: is the parameter of propensity to consume for the families. A high beta implies in more consumption of the family and less money saved for the real estate market.
\item \texttt{FPM\_DISTRIBUTION}: FPM is a municipal transfer tax in Brazil. FPM as 
`true', the default, means that taxes follow FPM distribution rules. 
\item \texttt{HOUSE\_VACANCY}: is the parameter used when generating houses that determines the amount of houses that will exceed the number of families in a given simulation configuration. 
\item \texttt{LABOR\_MARKET}: is the frequency the firm makes decisions on the labor market. 
\item \texttt{MARKUP}: is the percentage the firms use when they decide to raise prices. 
\item \texttt{MEMBERS\_PER\_FAMILY}: is also a parameter used to construct the agents. 
\item \texttt{PCT\_DISTANCE\_HIRING}: is the percentage of the sample of candidates in a given month that will be selected via the distance to the firm criteria. The remaining of the sample is chosen by qualification. 
\item \texttt{PERCENTAGE\_ACTUAL\_POP}: is the size of actual population to be included in the model. 
\item \texttt{PERCENTAGE\_CHECK\_NEW\_LOCATION}: is the sample of families at each given month that decide to enter the real estate market.
\item  \texttt{PROCESSING\_ACPS}: is the choice of ACP to be simulated.
\item \texttt{SIZE\_MARKET}: is the number of firms that families check when deciding on where to make purchases.
\item \texttt{STICKY\_PRICES}: is the factor that controls the frequency in which the firms evaluates whether to raise prices. 
\item \texttt{TAXES\_STRUCTURE}: those are approximations to the percentage of taxes that are distributed or transferred. 
\item \texttt{TAXES}: five taxes rates can be adjusted. 
\item \texttt{WAGE\_IGNORE\_UNEMPLOYMENT}: is the rule that tells the firms to either consider unemployment as a factor when deciding on wages (true) or not (false).
\end{enumerate}
\item The sensitivity params box is the place where you insert which parameters to test.
\item Run config: is the box related to the saving, plotting and printing of information from the simulations. \texttt{SAVE\_DATA} can include extra saving of details of `agents', `grave', `house', `family' and `house', which includes `firms'. 
\end{enumerate}	

\section{Validation and Illustration}
As a dynamic platform, PolicySpace is capable of replicating and maintaining the heterogeneity urban configuration of Brazilian metropolis as read from initial data. That means poorer, sparser housing and less dynamic firms in the outskirts and livelier, richer occupation in the core municipalities (Fig. a). Further, PolicySpace is capable of maintaining basic level macroeconomic indicators such as inflation and unemployment within windows of expected values despite significant variation on parameters and rules. 

\begin{figure}
\centering
\subfloat[Firms' profits for Belo Horizonte's ACP at the end of the simulation. Notice that there are a smaller number of firms outside the capital and that they also have smaller profits. Firms on the northwest municipality benefit from the distance to the main city and achieve average profit performance.]{{\includegraphics[width=5.6cm]{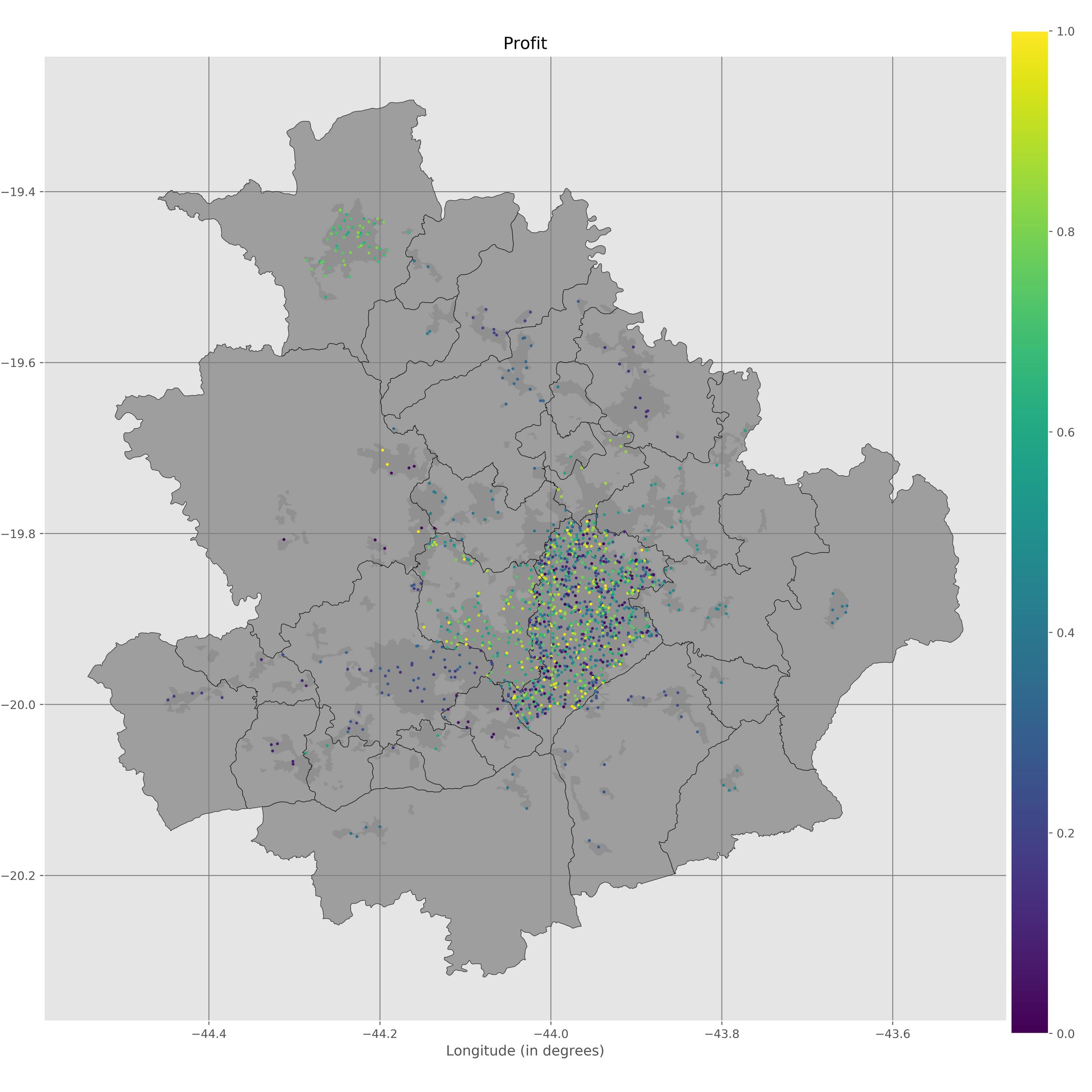}}}
\qquad
\subfloat[Total tax collection by ACPs: real data in blue and simulated data in green. This graph shows that the model is able to replicate closely the taxes collections of the set of ACPs modeled. A Kolmogorov-Smirnov test does not reject that the samples come from different distributions.]{{\includegraphics[width=5.6cm]{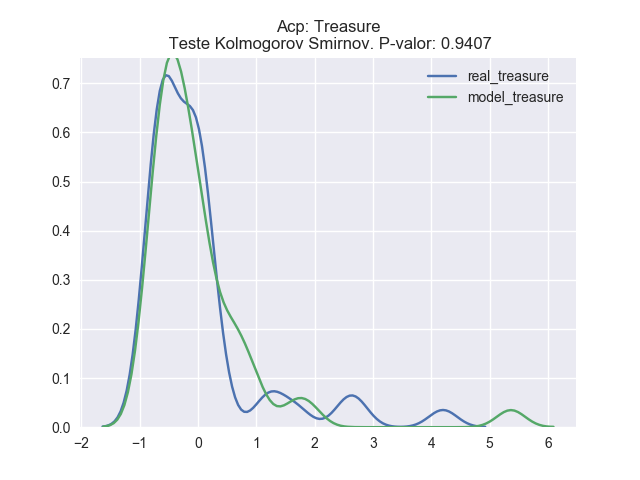}}}
\qquad
\subfloat[Evolution of Quality of Life indicator (QLI) for Bras\'{i}lia, 1\% of population. The graph shows true and false for \texttt{Alternative0}, given that FPM transfer rule is \texttt{False}. Results suggest that \texttt{Alternative0} as false (in blue), i.e. the municipalities are considered as one for fiscal distribution purposes, gives a positive boost to QLI which is equivalent to that obtained when the FPM transfer rule is in effect.]{{\includegraphics[width=5.6cm]{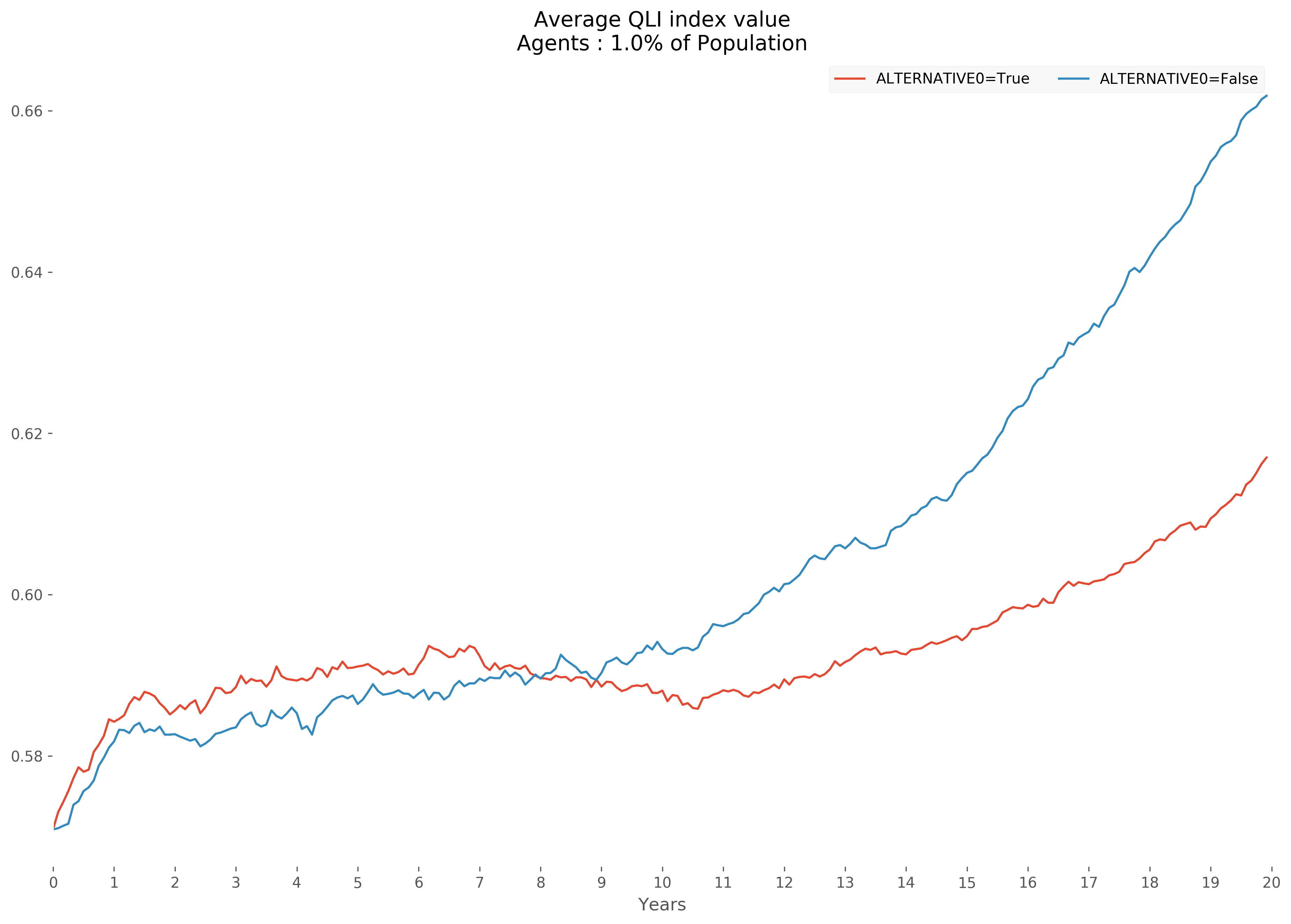}}}
\qquad
\subfloat[Results for all possible configurations of tax distribution for Rio de Janeiro's ACP, 2\% of population. When all distribution alternatives are tested together, the best result (in purple) happens when municipalities are considered as one for taxes purposes and the FPM transfer rule is in effect, although that is only marginally better than the observed \textit{status quo} (in red). The worst scenario would be not to have the FPM transfer rule and maintain the municipalities separated (in blue).]{{\includegraphics[width=5.6cm]{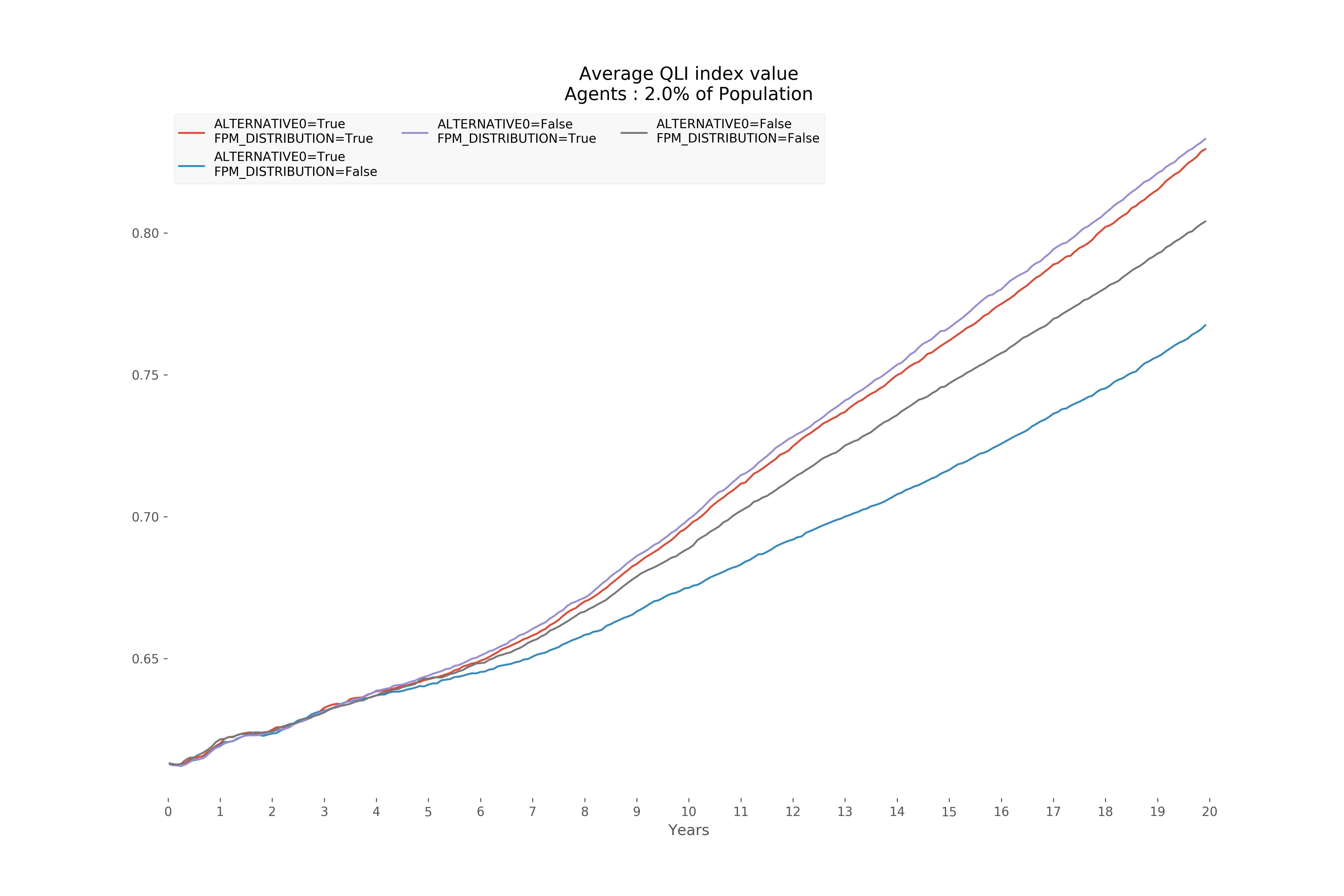}}}
\end{figure}

Considering that a key usability of PolicySpace is to allow for fiscal analysis among municipalities, main validation of the model comes from its capacity to replicate the aggregated behavior of taxes for the set of all ACPs. Simulating all 46 ACPs (333 municipalities) and using official data, PolicySpace maintains a proximal equivalence of magnitude with actual taxes (Fig. b). Actually, a Kolmogorov-Smirnov test \cite{kolmogorov_three_1965} does not reject the hypothesis that the two samples may have come from the same distribution. 

There are some differences, however. When comparing simulated and real data as percentage of total taxes, we see that:
\begin{itemize}
\item Property taxes are more concentrated in actual cases, whereas the simulated results are slightly more dispersed.
\item Transactions taxes on the other hand are a bit more concentrated in the simulated data when compared to actual data.
\end{itemize}

The illustration of taxes analysis (see Fig. c and Fig. d) clarifies the relevance of the two tax distribution schemes tests. First, it is clear that the existing FPM rule should be enforced as it helps increase the Quality of Life Indicator (QLI) in all scenarios. The test of the municipalities been together (\texttt{Alternative0:False}) or separate (\texttt{Alternative0:True}) for tax purposes, however, showed that when FPM is in place, the gain of the fusion of the municipalities is marginal. On the other hand, if there were no FPM in place, having the municipalities together would have an equivalent effect to the FPM transfer. All in all, the best theoretical and simulated result, which is consistent across different ACPs, is the one in which both the FPM transfer is present and the municipalities are considered as one, for tax purposes.

\section{Concluding considerations and further development}	
This paper has briefly introduced PolicySpace, an \textit{ex-ante} Public Policy analysis tool, demonstrated its main parameters and usage and illustrated the model with an application of alternative taxes distribution testing. However, part of the appeal of PolicySpace is that other modelers and researchers may use the model and change, adapt or evolve it for their own interests and research questions. Tentatively, we can think of at least some suggestions: 

\begin{itemize}
\item Further detailing the firms, providing them with differentiated sectors and products, testing decision-making strategies.
\item Making full use of the transversal and economically integrated commuting demand embedded in the model, considering that families are mobile and geocoded and that workers move employment among firms and that commuting is already computed within the model.
\item Make full use of all the data generated, especially real estate housing data, agents, families, firms, population dynamics and municipalities.
\item And, deeper testing and empirically detailing of taxes analysis and trickle effects within municipalities.
\end{itemize} 
\bibliography{policyspace}

\begin{thebibliography}{10}

\bibitem{abar_agent_2017}
{\sc Abar, S., Theodoropoulos, G.~K., Lemarinier, P., and O’Hare, G. M.~P.}
\newblock Agent based modelling and simulation tools: A review of the
  state-of-art software.
\newblock 13--33.

\bibitem{ahrend_what_2014}
{\sc Ahrend, R., Farchy, E., Kaplanis, I., and Lembcke, A.~C.}
\newblock What makes cities more productive? evidence on the role of urban
  governance from five {OECD} countries.
\newblock 33.

\bibitem{baptista_macroprudential_2016}
{\sc Baptista, R., Farmer, J.~D., Hinterschweiger, M., Low, K., Tang, D., and
  Uluc, A.}
\newblock Macroprudential policy in an agent-based model of the {UK} housing
  market.

\bibitem{batty_generic_2012}
{\sc Batty, M.}
\newblock A generic framework for computational spatial modelling.
\newblock In {\em Agent-based models of geographical systems}. Springer,
  pp.~19--50.

\bibitem{boero_agent-based_2015}
{\sc Boero, R., Morini, M., Sonnesa, M., and Terna, P.}
\newblock {\em Agent-based models of the economy: from theories to
  applications}.
\newblock Palgrave Macmillan.

\bibitem{colander_complexity_2014}
{\sc Colander, D., and Kupers, R.}
\newblock {\em Complexity and the Art of Public Policy: Solving Society's
  Problems from the Bottom Up}.
\newblock Princeton University Press.

\bibitem{dawid_agent-based_2014}
{\sc Dawid, H., Gemkow, S., Harting, P., Van~der Hoog, S., and Neugart, M.}
\newblock Agent-based macroeconomic modeling and policy analysis: the eurace@
  unibi model.

\bibitem{dosi_schumpeter_2010}
{\sc Dosi, G., Fagiolo, G., and Roventini, A.}
\newblock Schumpeter meeting keynes: A policy-friendly model of endogenous
  growth and business cycles.
\newblock 1748--1767.

\bibitem{edmonds_simulating_2013}
{\sc Edmonds, B., and Meyer, R.}
\newblock {\em Simulating Social Complexity: A Handbook}, 2013 edition~ed.
\newblock Springer.

\bibitem{epstein_growing_1996}
{\sc Epstein, J.~M., and Axtell, R.}
\newblock {\em Growing artificial societies: social science from the bottom
  up}.
\newblock Brookings/{MIT} Press.

\bibitem{erath_large-scale_2012}
{\sc Erath, A., Fourie, P., Van~Eggermond, M., Ord\'{o}\~{n}ez, S., Chakirov,
  A., and Axhausen, K.}
\newblock Large-scale agent-based transport demand model for singapore.
\newblock In {\em 13th International Conference on Travel Behaviour Research
  ({IATBR}). Toronto: International Association for Travel Behaviour Research}.

\bibitem{furtado_policyspace:_2018}
{\sc Furtado, B.~A.}
\newblock {\em {PolicySpace}: agent-based modeling}.
\newblock {IPEA}.

\bibitem{furtado_simple_2016}
{\sc Furtado, B.~A., and Eberhardt, I. D.~R.}
\newblock A simple agent-based spatial model of the economy: tools for policy.
\newblock 12.

\bibitem{furtado_fatos_2013}
{\sc Furtado, B.~A., Mation, L., and Monasterio, L.}
\newblock Fatos estilizados das finan\c{c}as p\'{u}blicas municipais
  metropolitanas brasileiras entre 2000-2010.
\newblock In {\em Territ\'{o}rio metropolitano, pol\'{i}ticas municipais}.
  Bernardo Alves Furtado; Cleandro Krause; Karla Fran\c{c}a, pp.~291--312.

\bibitem{gaffeo_adaptive_2008}
{\sc Gaffeo, E., Gatti, D.~D., Desiderio, S., and Gallegati, M.}
\newblock Adaptive microfoundations for emergent macroeconomics.
\newblock 441--463.

\bibitem{gasparini_transferencias_2011}
{\sc Gasparini, C.~E., and Miranda, R.~B.}
\newblock Transfer\^{e}ncias, equidade e efici\^{e}ncia municipal no brasil.

\bibitem{geyer_handbook_2015}
{\sc Geyer, R., and Cairney, P.}
\newblock {\em Handbook on complexity and public policy}.
\newblock Handbooks of Research on Public Policy series. Edward Elgar
  Publishing.

\bibitem{grimm_odd_2010}
{\sc Grimm, V., Berger, U., {DeAngelis}, D.~L., Polhill, J.~G., Giske, J., and
  Railsback, S.~F.}
\newblock The {ODD} protocol: a review and first update.
\newblock 2760--2768.

\bibitem{hamill_agent-based_2016}
{\sc Hamill, L., and Gilbert, N.}
\newblock {\em Agent-Based modelling in economics}.
\newblock Wiley.

\bibitem{helbing_social_2012}
{\sc Helbing, D.}
\newblock {\em Social Self-Organization: Agent-Based Simulations and
  Experiments to Study Emergent Social Behavior}.
\newblock Understanding Complex Systems. Springer.

\bibitem{heppenstall_agent-based_2012}
{\sc Heppenstall, A.~J., Crooks, A.~T., See, L.~M., and Batty, M.}
\newblock Agent-based models of geographical systems.
\newblock 746.

\bibitem{kolmogorov_three_1965}
{\sc Kolmogorov, A.~N.}
\newblock Three approaches to the quantitative definition of information'.
\newblock 3--11.

\bibitem{lengnick_agent-based_2013}
{\sc Lengnick, M.}
\newblock Agent-based macroeconomics: A baseline model.
\newblock 102--120.

\bibitem{neugart_agent-based_2012}
{\sc Neugart, M., and Richiardi, M.}
\newblock Agent-based models of the labor market.

\bibitem{oecd_applications_2009}
{\sc {OECD}}.
\newblock {\em Applications of complexity science for public policy: new tools
  for finding unanticipated consequences and unrealized opportunities}.
\newblock {OECD}.

\bibitem{schelling_process_1972}
{\sc Schelling, T.~C.}
\newblock A process of residential segregation: neighborhood tipping.
\newblock 174.

\bibitem{seppecher_what_2017}
{\sc Seppecher, P., Salle, I., and Lavoie, M.}
\newblock What drives markups? evolutionary pricing in an agent-based
  stock-flow consistent macroeconomic model.

\bibitem{tabak_topological_2009}
{\sc Tabak, B.~M., Cajueiro, D.~O., and Serra, T.~R.}
\newblock Topological properties of bank networks: the case of brazil.
\newblock 1121--1143.

\bibitem{wilensky_introduction_2015}
{\sc Wilensky, U., and Rand, W.}
\newblock {\em An introduction to Agent-Based Modeling}.
\newblock The {MIT} Press.

\end{thebibliography}
\bibliographystyle{acm}
\end{document}